\documentclass[12pt,twoside]{article}
\usepackage{xrb2007}
\begin{document}

\session{Population Synthesis}

\shortauthor{Fregeau}
\shorttitle{X-Ray Binaries and the Dynamical States of Globular Clusters}

\title{X-Ray Binaries and the Dynamical States of Globular Clusters}
\author{John M. Fregeau}
\affil{Chandra Fellow; Department of Physics and Astronomy, Northwestern University, Evanston, IL 60208}

\begin{abstract}
We summarize and discuss recent work (Fregeau 2007) that presents the confluence of three results suggesting
that most Galactic globular clusters are still in the process of core contraction,
and have not yet reached the thermal equilibrium
phase driven by binary scattering interactions: that
1) the three clusters that appear to be overabundant in X-ray binaries per unit
encounter frequency are
observationally classified as ``core-collapsed,'' 2) recent numerical simulations
of cluster evolution with primordial binaries show that structural parameters
of clusters in the binary-burning phase agree only with ``core-collapsed''
clusters, and 3) a cluster in the binary-burning phase for the last few Gyr should 
have $\sim 5$ times more dynamically formed X-ray sources than if it 
were in the core contraction phase for the same time.
\end{abstract}

\section{Introduction}
This proceedings article briefly summarizes my contributed talk at the ``Population Explosion''
meeting, which itself was a brief summary of the work described in \citet{fregeau07}.  I
have taken the opportunity here to include supplemental material, but first refer
the reader to the detailed, complete discussion in \citet{fregeau07}.

\subsection{X-Ray Sources and Cluster Dynamics}
Although X-ray binaries have been known to be overabundant in Galactic globular clusters 
(by unit mass relative to the disk) for over 30 years
\citep{1975ApJ...199L.143C,1975Natur.253..698K}, it is only recently that clear evidence 
for their dynamical origin has revealed itself.  \citet{2003ApJ...591L.131P} found
that the number of X-ray sources with $L_X > 4 \times 10^{30}\,{\rm erg}/{\rm s}$, $N_X$, 
in a cluster
follows a clear, nearly linear correlation with the encounter frequency $\Gamma$,
a measure of the dynamical interaction rate in the cluster.  There are a few clear
outliers to this trend, however.  Terzan 1 is overabundant in $N_X$ by a factor of 
$\sim 20$, NGC 6397 by a factor of $\sim 5$, and NGC 7099 by a factor of $\sim 2$
\citep{2006MNRAS.369..407C,2007ApJ...657..286L}.
The common thread among these clusters is that they are all classified observationally
as ``core-collapsed,'' while the other clusters in the sample are not.  A cluster is
observationally termed core-collapsed if its surface brightness profile is consistent
with a cusp at the limit of observational resolution.  Since this observational
classification is linked with the cluster evolutionary state (as described
below), it appears that cluster evolution complicates the $N_X$--$\Gamma$ correlation.

\subsection{Understanding Cluster Core Radii}
The evolution of a globular cluster, being a bound self-gravitating system,
is very similar to that of a star, and comprises three main phases.
The cluster first ``core contracts'' on a relaxation timescale.  Once
the core density becomes large enough for binary scattering interactions
to begin generating energy, the cluster settles into the ``binary-burning''
phase, in which the cluster's core properties remain roughly constant with time.  
Once the binary population is exhausted in the core, it collapses
via the gravothermal instability, leading to extremely high central densities, followed
by a series of gravothermal oscillations in which the core expands and contracts
repeatedly.  (For a graphical representation of the three main phases of 
cluster evolution, see Figure 1 of \citet{1991ApJ...370..567G}, Figure 5 of \citet{2003ApJ...593..772F},
or Figure 29.1 of \citet{2003gmbp.book.....H}.)
Recent comparison of star cluster evolution simulations with observations of cluster
structural parameters have shown that the clusters observationally
classified as core-collapsed are in fact most likely in the binary-burning
phase, while the rest are most likely still in the core-contraction phase
\citep{2006MNRAS.368..677H,2007ApJ...658.1047F}.

\subsection{A Refined $\Gamma$}
First introduced by \citet{1987IAUS..125..187V}, the encounter frequency
$\Gamma$ is an estimate of the {\em current} dynamical interaction rate
in the cluster, which is assumed to be proportional to the current
number of observable X-ray sources.  Typically, it is approximated
as \begin{equation}\label{eq:standardgamma}
  \Gamma \equiv \frac{dN_{\rm int}}{dt} \propto \rho_c^2 r_c^3 / v_\sigma \, ,
\end{equation}
where $r_c$ is the cluster core radius, $\rho_c$ is the core mass density, 
and $v_\sigma$ is the core velocity dispersion.  More accurate approximations of 
the interaction rate have been used, including numerical integrals over 
cluster models \citep{2003ApJ...591L.131P}, but all are estimates of 
the {\em current} interaction rate.  Recent work has shown that 
dynamically formed X-ray binaries have finite detectable lifetimes 
($\sim 10^5$ to $\sim 10^9$ yr, depending on binary type), and furthermore
that there is a lag time of several Gyr between dynamical interaction
and the binary turning on as an X-ray source \citep{2006MNRAS.372.1043I,2007arXiv0706.4096I}.
Since the binary interaction rate is a function of the cluster properties (density,
velocity dispersion, etc.), it is clear that the X-ray binaries we see in clusters today were formed
several Gyr ago, and thus encode the recent cluster evolution history in their
populations.  We thus adopt a refined version of $\Gamma$ which encodes this history
by integrating the interaction rate over the dynamical evolution of the cluster.  
The resulting quantity is the total number of strong dynamical interactions, 
which should be roughly proportional to the currently observable
number of X-ray binaries.  Plugging in the different phases of cluster evolution 
will yield different estimates for the number of sources, and possibly allow one 
to differentiate among the different phases by comparing with the observed
number of sources.  Using standard assumptions, and adopting some results from 
recent $N$-body simulations for the evolution of $r_c$ with time, one finds
the ratio of the number of interactions for a cluster in the binary-burning phase
to one in the core contraction phase is
\begin{equation}
  \frac{N_{\rm int,bb}}{N_{\rm int,cc}} = \frac{t_x}{t_0}
  \frac{2.835}{\left(\frac{t_0}{t_0+9t_\ell}\right)^{0.315}-
    \left(\frac{t_0}{t_0+9t_\ell+9t_x}\right)^{0.315}} \, ,
\end{equation}
where $t_x$ is the X-ray source lifetime, $t_\ell$ is the lag time between
strong dynamical interaction and X-ray turn-on, and $t_0$ is the current
cluster age.  (The gravothermal oscillation phase is excluded from the 
discussion since in this phase
the core binary fraction would be much smaller than what is observed.)
This expression has a minimum of $2.0$ and a maximum of $17.8$ in the range
$t_x=10^{-4}$--$3\,{\rm Gyr}$,
$t_\ell=1$--$10\,{\rm Gyr}$, for $t_0=13\,{\rm Gyr}$.
For the canonical values of $t_x=1\,{\rm Gyr}$
and $t_\ell=3\,{\rm Gyr}$ with $t_0=13\,{\rm Gyr}$, the value is $5.0$.  Since the 
number of X-ray sources should scale roughly linearly with the number of interactions,
this suggests that if a cluster is in the binary-burning phase (and has been for a time
$t_\ell+t_x$ to the present), it should have $\sim 5$ times
as many X-ray sources than it would if it were in the core contraction phase.
The three observationally core-collapsed clusters in the sample previously mentioned
are overabundant in $N_X$ by factors compatible with the range of values allowed
by this expression, suggesting that the observationally core-collapsed clusters are 
indeed in the binary-burning phase, while the rest are still in the process of core contraction.

\subsection{Discussion}
Since only $\sim 20$\% of Galactic globular clusters are observationally
classified as core-collapsed, the conclusion that seems strongly suggested
is that most clusters ($\sim 80$\%) are currently still in the core contraction phase, 
while those that are core-collapsed are in the binary
burning phase.  This goes counter to the widely held belief that most clusters 
are currently in the binary burning phase, and complicates the many existing studies that have assumed
cluster core properties that are constant with time, including predictions of blue
straggler populations, tidal-capture binaries, and the evolution of the core
binary fraction \citep[e.g.,][]{2004ApJ...605L..29M,1994ApJ...423..274D,2005MNRAS.358..572I}.  
Similarly, this result does away with the need for alternate
energy sources in cluster cores to explain currently observed core radii,
including intermediate-mass black holes in most clusters, mass loss
from stellar mergers, and ongoing mass segregation
\citep{2006astro.ph.12040T,chatterjeeposter,2004ApJ...608L..25M}.

\section{Meta-Discussion}
A few interesting aspects of this analysis did not fit into the original {\em letter}.
It has been pointed out that an important but neglected factor going into the interaction
rate is the neutron star retention fraction, since it depends on the cluster
escape speed, and since it is neutron star $+$ stellar
binary interactions that predominantly lead to X-ray binaries (Rasio, private communication).
Since there is no reason to expect that the cluster escape speed at the time of
neutron star formation does not vary from cluster to cluster, the correlation 
between dynamical interaction rate and $N_X$ should
be blurred significantly by this effect.  We note that the analysis
in \citet{fregeau07} avoids this issue entirely by comparing a cluster only with
a version of itself in a different evolutionary phase.

Another issue is that there is apparently a significant contribution to the X-ray
binary population from primordial binaries in clusters with small encounter
frequencies.  (This is likely the reason for the sub-linear dependence of
$N_X$ on $\Gamma$ in \citet{2003ApJ...591L.131P}.)  This has given rise to
the use of the interaction rate per unit cluster mass, $\gamma \equiv \Gamma/M_{\rm clus}$,
to help isolate the dynamically-formed population in the analysis.
The use of $\gamma$ has, for example, shown that a significant fraction of cataclysmic variables
are dynamically formed \citep{2006ApJ...646L.143P}.  The results of \citet{fregeau07}
could be further tested in the future with the more sensitive $\gamma$ 
as the library of deep X-ray observations of globular clusters grows.

Finally, we mention that in the near future we plan to test our simple overabundance
prediction by using the detailed numerical models of \citet{2005MNRAS.358..572I}.
Although there are some hints from that paper that inputting a time-varying core does not noticeably
affect the final core binary fraction, we expect that the number of visible X-ray binaries,
while noisy, may be more noticeably affected.

\acknowledgments
JMF acknowledges support from Chandra theory grant TM6-7007X and Chandra 
Postdoctoral Fellowship Award PF7-80047.

\end{document}